\documentclass[aps,prd,10pt,onecolumn,nofootinbib]{revtex4-2}

\usepackage[T1]{fontenc}
\usepackage[utf8]{inputenc}
\usepackage{amsmath,amssymb,bm,mathtools}
\usepackage{graphicx}
\usepackage{hyperref}
\hypersetup{hidelinks,pdftitle={Light-front diagnostics in the 't Hooft model: I. Wave functions, EMT decomposition, and the diagonal GPD overlap},pdfauthor={Arkadiy I. Syamtomov}}
\usepackage{booktabs}
\usepackage{float}
\usepackage[section]{placeins}

\newcommand{\mtilde}{\widetilde m}

\newcommand{\MD}{\mathcal M_{2,n}^{\mathrm D}}
\newcommand{\dd}{\mathrm d}

\begin{document}

\title{Light-front diagnostics in the 't~Hooft model: I. Wave functions, EMT decomposition, and the diagonal GPD overlap}

\author{Arkadiy I. Syamtomov}
\affiliation{Bogolyubov Institute for Theoretical Physics\\National Academy of Sciences of Ukraine\\14-B Metrolohichna Street, 03143 Kyiv, Ukraine}

\begin{abstract}\noindent
I examine how the longitudinal light-front wave function of a meson encodes forward energy--momentum tensor (EMT) structure and the diagonal part of an off-forward generalized parton distribution in the large-$N_c$ 't~Hooft model.  Light--light, equal-mass reference, heavy--light, and heavy--heavy systems are compared through their momentum distributions, differential entropy, bilocal Coulomb kernel, and forward mass-squared decomposition.  An independent sine-basis calculation confirms the light--light spectrum despite the slow convergence of the sine representation.  For equal constituent masses, the second moment obtained from the exact diagonal overlap has no term linear in the asymmetric skewness variable $b$, while its regular $b^2$ and $b^3$ coefficients are equal and determined by a universal kinematic contribution and a weighted wave-function gradient norm.  At the equal-mass reference point $\mtilde^{\,2}=1$, corresponding to $\beta=1/2$, the boundary expansion becomes resonant and generates a $b^4\ln^2(1/b)$ nonanalyticity, identifying the precise limitation of the diagonal two-body overlap.  The companion Part~II analysis constructs the ERBL completion required to cancel this support-dependent nonanalyticity and restore the analyticity of the local EMT moment.
\end{abstract}

\maketitle

\section{Introduction}
\label{sec:intro}

The large-$N_c$ 't~Hooft model was introduced as a tractable theory of confined mesons in two-dimensional QCD~\cite{tHooft:1974}.  Its light-front formulation makes it particularly useful for controlled studies of hadron structure.  At leading order in $1/N_c$, a meson is described by a single $q\bar q$ light-front wave function depending on the boost-invariant longitudinal momentum fraction $x$.  This amplitude determines the bound-state spectrum, forward parton distributions, forward local matrix elements, and the diagonal two-body contributions to off-forward bilocal matrix elements, while the confining instantaneous interaction remains explicit in the Hamiltonian.  The model therefore provides a direct setting in which wave-function structure and energy--momentum tensor (EMT) matrix elements can be compared~\cite{Brodsky:1998}.

Several ingredients needed for such a comparison are already known.  The original bound-state equation determines the meson spectrum and light-front wave functions~\cite{tHooft:1974}.  Forward light-front sum rules relate the invariant mass squared to scalar-charge and Coulomb-energy matrix elements, and the corresponding EMT decomposition has been analyzed in Refs.~\cite{Ji:2021,Freese:2023}.  Off-forward parton distributions in the model were derived by Burkardt~\cite{Burkardt:2000}, and the complete light-cone GPD, including its nonvalence region, was constructed more recently by Jia \textit{et al.}~\cite{Jia:2024}.  Quantum entanglement associated with momentum-fraction partitions and rapidity evolution, as well as spatial entanglement in two-dimensional QCD, has also been studied~\cite{Liu:2022a,Liu:2022b}.  These entanglement entropies are conceptually distinct from the differential Shannon entropy used below as a simple localization diagnostic.

This is the first of two companion papers organized according to the dynamical sectors required for the analysis.  Part I asks what can be extracted from the leading two-body wave function and its diagonal GPD overlap: the longitudinal structure, forward EMT decomposition, and the near-forward boundary obstruction.  The companion Part II paper treats the nonvalence ERBL sector, its intermediate-meson representation, and the restoration of the analytic local EMT form factor.  This division separates two-body information from nonvalence completion; it is not simply a division between forward and nonforward kinematics.

The detailed wave-function shape, the forward decomposition of the meson invariant mass squared, the bilocal Coulomb pattern, and the diagonal part of the GPD moment have generally been studied separately, leaving their correlated state and mass dependence only partly explicit.  The purpose of the present paper is to place these quantities on the same footing and to identify precisely which EMT information is already fixed by the leading two-body wave function.

I first compare four representative mass regimes and determine how their longitudinal momentum distributions change with excitation.  I then connect those changes to the explicit quark-mass and Coulomb-interaction contributions to $M_n^2$.  Finally, I analyze the diagonal GPD second moment near the forward point and show how the near-boundary expansion of the wave function controls its analytic structure.

The model has no transverse momentum transfer and hence cannot define intrinsic impact-parameter densities, pressure profiles, or shear distributions.  Those notions require genuine $3+1$-dimensional kinematics~\cite{Diehl:2003,Polyakov:2018}.  The 't~Hooft model instead isolates the longitudinal relations among confinement, light-front wave functions, GPD support, and EMT matrix elements without introducing additional transverse modeling.

\section{Meson wave functions and longitudinal structure}
\label{sec:wf}

To relate the internal structure of a meson to its EMT, the bound state must first be determined with sufficient control near $x=0$ and $x=1$.  In the large-$N_c$ theory this information is contained in one light-front wave function $\phi_n(x)$, where $x$ is the fraction of the meson longitudinal momentum carried by constituent 1; constituent 2 carries $1-x$.  Accordingly, $m_1$ multiplies the $1/x$ term and $m_2$ the $1/(1-x)$ term below.

I use the conventional scale
\begin{equation}
 \lambda=\frac{g^2N_c}{\pi},
 \qquad
 \mtilde_i^2=\frac{m_i^2}{\lambda},
 \qquad
 M_n^2\equiv\frac{M_{n,\mathrm{phys}}^2}{\lambda}.
 \label{eq:units}
\end{equation}
Thus $M_n^2$ and all momentum transfers below are dimensionless unless stated otherwise; the tilde on the meson mass is suppressed.  In these units the 't~Hooft equation is~\cite{tHooft:1974}
\begin{equation}
 M_n^2\phi_n(x)=
 \left(\frac{\mtilde_1^2}{x}+\frac{\mtilde_2^2}{1-x}\right)\phi_n(x)
 -\mathrm P\!\int_0^1\!\dd y\,
 \frac{\phi_n(y)-\phi_n(x)}{(x-y)^2},
 \qquad x\in(0,1).
 \label{eq:thooft}
\end{equation}
The first term contains the explicit constituent masses, whereas the singular integral is the instantaneous confining interaction.  The subtraction removes the coincident-point singularity at $y=x$, while the remaining near-boundary balance with the mass terms dynamically fixes the exponents $\beta_1$ and $\beta_2$ in Eq.~\eqref{eq:betaindividual}.

Substituting $\phi_n(x)\sim x^{\beta_1}$ near $x=0$ and $\phi_n(x)\sim(1-x)^{\beta_2}$ near $x=1$ gives~\cite{tHooft:1974}
\begin{equation}
 \mtilde_i^2-1+\pi\beta_i\cot(\pi\beta_i)=0,
 \qquad
 \phi_n(x)\sim x^{\beta_1}(1-x)^{\beta_2}.
 \label{eq:betaindividual}
\end{equation}
The exponents $\beta_i$ are therefore determined by the quark masses rather than introduced as new parameters.  Because they are generally noninteger, a polynomial basis that ignores them converges slowly, especially for light quarks.  I instead factor out the mass-determined boundary behavior $x^{\beta_1}(1-x)^{\beta_2}$ and approximate only the remaining regular interior function.

Motivated by boundary-adapted basis approaches~\cite{Abe:2000} and by the documented slow small-mass convergence of representations that do not resolve the boundary powers~\cite{vanDeSande:1996}, I use the Jacobi realization
\begin{equation}
 \phi_n(x)=\sum_{k=0}^{N_B-1}c_k^{(n)}f_k(x),
 \qquad
 f_k(x)=x^{\beta_1}(1-x)^{\beta_2}
 P_k^{(2\beta_2,2\beta_1)}(2x-1).
 \label{eq:jacobibasis}
\end{equation}
To obtain the finite-dimensional bound-state equation, I insert Eq.~\eqref{eq:jacobibasis} into Eq.~\eqref{eq:thooft}, multiply by a test function $f_k(x)$, and integrate over $x$.  The explicit mass terms are projected directly.  For the singular interaction, exchanging $x$ and $y$ in one copy of the integral and averaging the two representations gives
\begin{equation*}
 \int_0^1\dd x\,f_k(x)\,\mathrm P\!\int_0^1\dd y\,
 \frac{f_l(x)-f_l(y)}{(x-y)^2}
 =\frac12\int_0^1\dd x\int_0^1\dd y\,
 \frac{[f_k(x)-f_k(y)][f_l(x)-f_l(y)]}{(x-y)^2}.
\end{equation*}
This symmetrization removes the apparent diagonal singularity and yields the matrix elements
\begin{align}
 H_{kl}={}&\int_0^1\dd x\,f_k(x)f_l(x)
 \left(\frac{\mtilde_1^2}{x}+\frac{\mtilde_2^2}{1-x}\right)
 \nonumber\\
 &+\frac12\int_0^1\dd x\int_0^1\dd y\,
 \frac{[f_k(x)-f_k(y)][f_l(x)-f_l(y)]}{(x-y)^2}.
 \label{eq:galerkinH}
\end{align}
Because the basis functions are not orthonormal under the ordinary inner product, the projection of the left-hand side $M_n^2\phi_n(x)$ introduces the overlap matrix.  The coefficients therefore satisfy the generalized eigenvalue problem
\begin{equation}
 \sum_l H_{kl}c_l^{(n)}
 =M_n^2\sum_l A_{kl}c_l^{(n)},
 \qquad
 A_{kl}=\int_0^1 f_k(x)f_l(x)\,\dd x.
 \label{eq:generalizedEigen}
\end{equation}
Finally, the physical orthonormality condition $\int_0^1\phi_m(x)\phi_n(x)\,\dd x=\delta_{mn}$ becomes, after inserting the basis expansion,
\begin{equation}
 \sum_{kl}c_k^{(m)}A_{kl}c_l^{(n)}=\delta_{mn}.
 \label{eq:Anorm}
\end{equation}
With the normalization chosen in Eq.~\eqref{eq:jacobibasis}, Jacobi orthogonality makes $A_{kl}$ diagonal, but the individual basis functions do not have unit norm, so $A_{kk}\neq1$.  Consequently, Eq.~\eqref{eq:generalizedEigen} is a generalized eigenvalue problem and the eigenvectors are normalized with the $A$-metric in Eq.~\eqref{eq:Anorm}.

For each mass choice I evaluate $H_{kl}$ and $A_{kl}$, solve Eq.~\eqref{eq:generalizedEigen}, order the eigenstates by $M_n^2$, and impose Eq.~\eqref{eq:Anorm}.  The singular interaction is always evaluated through the symmetric quadratic form in Eq.~\eqref{eq:galerkinH}.  The change of variable $x=(1-\cos\theta)/2$ increases the resolution near both boundaries, and the basis dimension and integration order are varied independently.  The six lowest eigenvalues, $n=0,\ldots,5$, are stable at the level required below.  Because the light--light family has the smallest boundary exponent and converges most slowly, its spectrum is also checked with an independent sine-basis variational calculation in Appendix~\ref{app:numerics}.

For a normalized state, the longitudinal momentum-fraction probability density, its moments, and its variance are
\begin{equation}
 q_n(x)=\phi_n^2(x),
 \qquad
 \int_0^1 q_n(x)\,\dd x=1,
 \qquad
 \langle x^r\rangle_n=\int_0^1x^r q_n(x)\,\dd x,
 \qquad
 \sigma_{x,n}^2=\langle x^2\rangle_n-\langle x\rangle_n^2.
 \label{eq:qn_moments}
\end{equation}
The variance $\sigma_{x,n}^2$ measures the mean-square displacement of the longitudinal momentum fraction from its mean, but it does not retain the full shape of an excited-state distribution.  Internal nodes of $\phi_n(x)$ divide $q_n(x)$ into several probability lobes, and different multi-lobed distributions can have the same second central moment.

To characterize how broadly the longitudinal momentum fraction is distributed, I compare $q_n(x)$ with the uniform probability density
\begin{equation*}
 q_{\mathrm{unif}}(x)=1,
 \qquad
 0\leq x\leq1,
 \qquad
 \int_0^1 q_{\mathrm{unif}}(x)\,\dd x=1.
\end{equation*}
This density is maximally delocalized on the fixed momentum-fraction range and therefore provides a natural reference for the shape of $q_n(x)$.  The degree of localization is quantified by the differential Shannon entropy,
\begin{equation}
 S_n^x=-\int_0^1q_n(x)\ln q_n(x)\,\dd x
      =-D_{\mathrm{KL}}\!\left(q_n\Vert q_{\mathrm{unif}}\right).
 \label{eq:Sn}
\end{equation}
Thus $S_n^x$ is the negative relative entropy with respect to the normalized uniform reference.  Differential entropy is not invariant under a nonlinear change of variable.  Here every state is evaluated in the same boost-invariant momentum fraction $x$, with $0\leq x\leq1$, which makes the entropy values directly comparable.  The entropy measures the overall concentration of the full distribution, although it does not locate individual nodes.  It is not a thermodynamic entropy or a von Neumann entanglement entropy of the type studied in Refs.~\cite{Liu:2022a,Liu:2022b}.

For the uniform reference distribution,
\begin{equation}
 \langle x\rangle_{\mathrm{unif}}=\frac12,
 \qquad
 \sigma_{x,\mathrm{unif}}^2
 =\int_0^1\left(x-\frac12\right)^2\dd x
 =\frac1{12},
 \qquad
 S_{\mathrm{unif}}=0.
 \label{eq:uniformvariance}
\end{equation}
The mean value $1/2$ reflects symmetry about the center of the momentum-fraction range, while $1/12$ is the variance of a distribution spread uniformly over that entire range.  The entropy value $S_{\mathrm{unif}}=0$ is the maximal value for a normalized density with $0\leq x\leq1$.  These numbers therefore provide separate reference points for the centroid, width, and localization of the meson distributions.  A variance larger than $1/12$ means that the mean-square displacement from the mean is larger than for the uniform distribution; it does not by itself imply a uniformly broader shape.

To separate the effects of the overall mass scale, mass asymmetry, and excitation level, I choose four dimensionless squared-mass pairs,
\begin{equation}
 (\mtilde_1^2,\mtilde_2^2)=
 (0.04,0.04),\quad(1,1),\quad(0.04,4),\quad(16,16).
 \label{eq:massfamilies}
\end{equation}
These are controlled benchmarks chosen here, not assignments to physical mesons.  The value $0.04$ gives a small boundary exponent and a broad relativistic state.  The equal-mass point $\mtilde^2=1$ gives $\beta=1/2$ and is special because the subleading boundary expansion becomes logarithmic.  The pair $(0.04,4)$ probes strongly asymmetric heavy--light momentum sharing, whereas $(16,16)$ produces a localized heavy--heavy state whose ground-state mass is dominated by the explicit mass term.  Together the four families separate mass scale, asymmetry, and excitation effects.  The first four eigenvalues are collected in Table~\ref{tab:spectrum}; the independent light-sector check is discussed in Appendix~\ref{app:numerics}.

\begin{table}[!htbp]
\caption{Dimensionless squared masses $M_n^2$ for the first four eigenstates.  Parentheses give the numerical uncertainty in the final displayed digits.}
\label{tab:spectrum}
\begin{ruledtabular}
\begin{tabular}{lrrrr}
Family & $n=0$ & $n=1$ & $n=2$ & $n=3$ \\
\hline
light--light & $0.8694(38)$ & $7.660(14)$ & $16.197(27)$ & $25.252(47)$ \\
equal-mass reference & $7.2745(19)$ & $17.309(11)$ & $27.123(28)$ & $37.021(51)$ \\
heavy--light & $8.5371(64)$ & $19.587(17)$ & $30.224(34)$ & $40.652(59)$ \\
heavy--heavy & $74.8065(81)$ & $94.376(35)$ & $109.457(75)$ & $123.82(13)$ \\
\end{tabular}
\end{ruledtabular}
\end{table}
The quoted uncertainties are conservative estimates from the quadrature extrapolation.  The independent light-sector calculation in Appendix~\ref{app:numerics} provides an additional sine-basis check, while recent analytic studies give complementary information on the equal- and unequal-mass spectra, including the heavy--light limit~\cite{Litvinov:2025,Artemev:2025}.

\begin{figure}[!htbp]
 \centering
 \includegraphics[width=0.96\columnwidth]{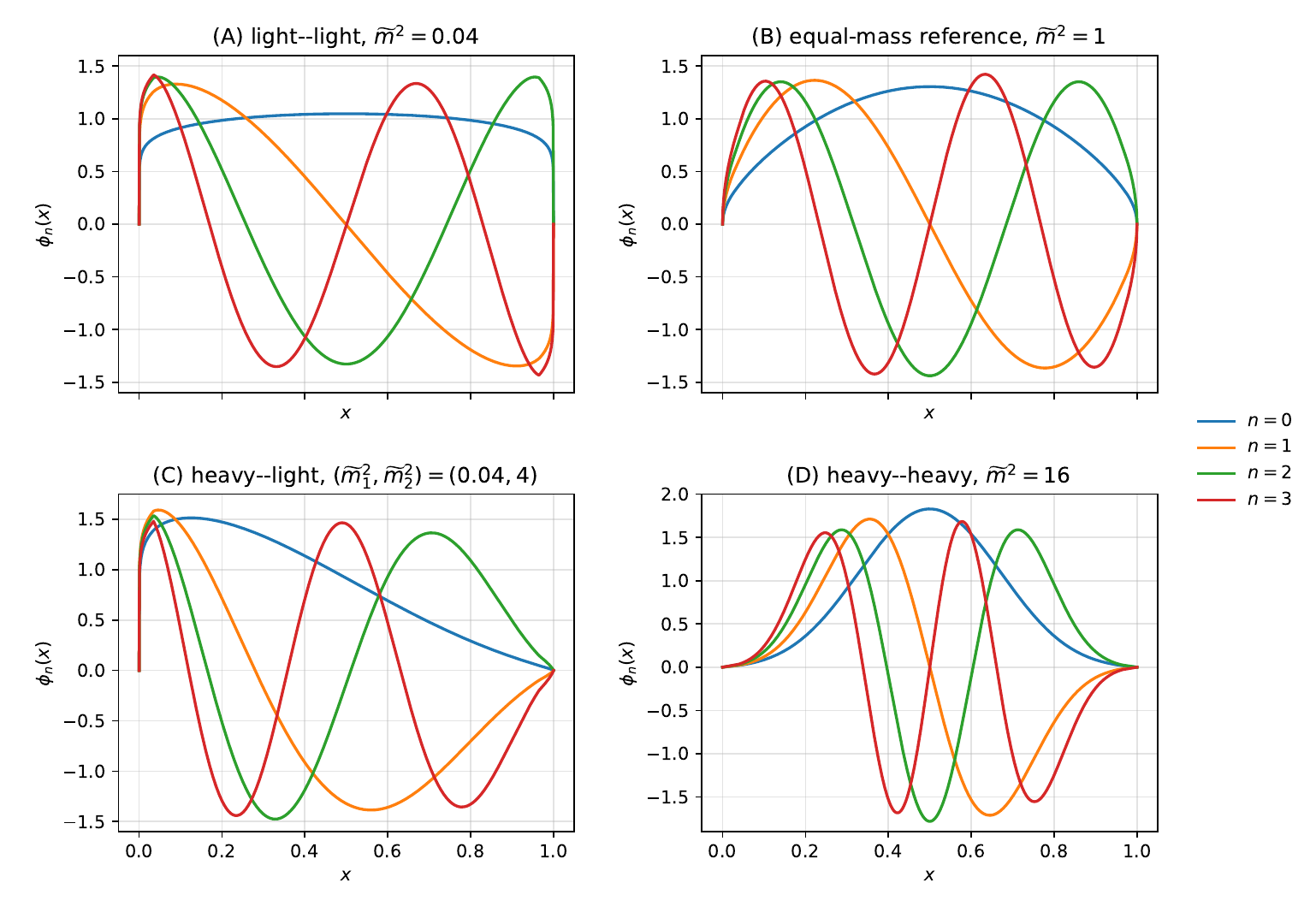}
 \caption{Light-front wave functions $\phi_n(x)$ for the four mass families and $n=0,1,2,3$.  The exact endpoint values $\phi_n(0)=\phi_n(1)=0$ are included.}
 \label{fig:wavefunctions}
\end{figure}

Wave functions and momentum distributions in the 't~Hooft model have been calculated in the original solution and in later studies of partonic structure and GPDs~\cite{tHooft:1974,Burkardt:2000,Ji:2021,Jia:2024}.  Fig.~\ref{fig:wavefunctions} shows the signed light-front amplitudes calculated here for the four mass choices in Eq.~\eqref{eq:massfamilies}.  The overall sign of each eigenfunction is conventional; its nodes and relative sign changes across the interval are the meaningful features.  For equal constituent masses, the equation is invariant under $x\leftrightarrow1-x$, so the eigenfunctions can be chosen to be symmetric or antisymmetric about $x=1/2$.  With increasing excitation level, the number of interior zeros increases, producing the expected alternating reflection parity and nodal pattern.  In the heavy--light family the wave function is asymmetric, reflecting momentum sharing in which the heavier constituent carries the larger fraction.  This comparison therefore makes visible both the mass-controlled boundary behavior and the nodal structure that is lost after squaring the amplitude.

\FloatBarrier

\begin{figure}[!htbp]
 \centering
 \includegraphics[width=0.96\columnwidth]{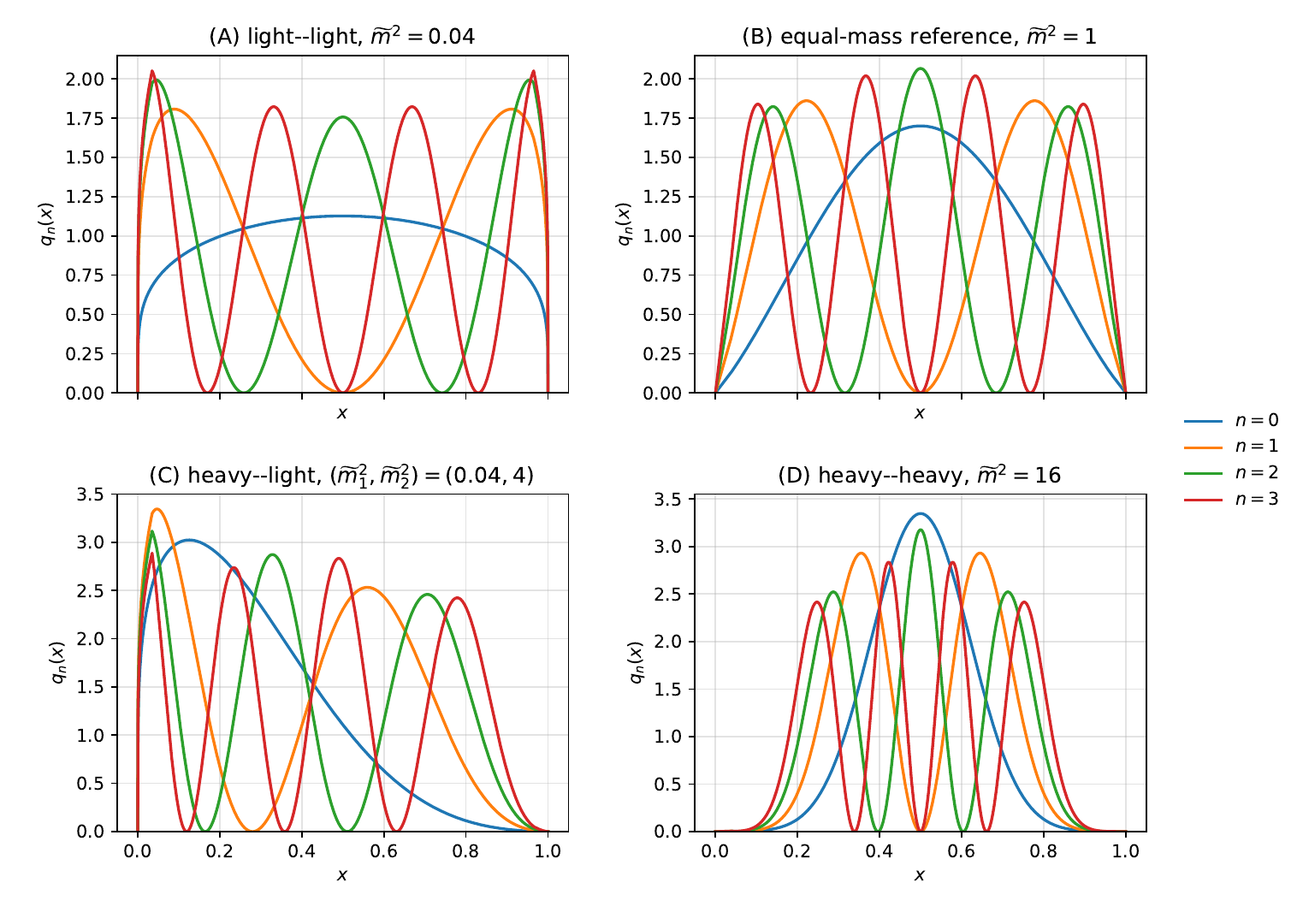}
 \caption{Longitudinal momentum distributions $q_n(x)=\phi_n^2(x)$ for the same states as in Fig.~\ref{fig:wavefunctions}.  The exact endpoint values $q_n(0)=q_n(1)=0$ are included.}
 \label{fig:pdfs}
\end{figure}

The corresponding momentum distributions $q_n(x)=\phi_n^2(x)$ are shown in Fig.~\ref{fig:pdfs}.  Squaring removes the sign changes while retaining the nodal zeros and the separation into probability lobes, so these curves are the appropriate objects for the moments in Eq.~\eqref{eq:qn_moments}.  The light--light ground state is broad, with $\sigma_{x,0}^2=0.0712$, while its first excitation places more probability away from the mean than the uniform reference.  The equal-mass excited states lie close to $1/12$, although their multi-lobed structure is clearly nonuniform.  The heavy--heavy ground state is strongly localized near $x=1/2$, with $\sigma_{x,0}^2=0.0139$, and remains below the uniform reference over the displayed states.  The heavy--light distributions are asymmetric at low excitation and move gradually toward more equal momentum sharing.

\FloatBarrier

\begin{figure}[!htbp]
 \centering
 \includegraphics[width=0.90\columnwidth]{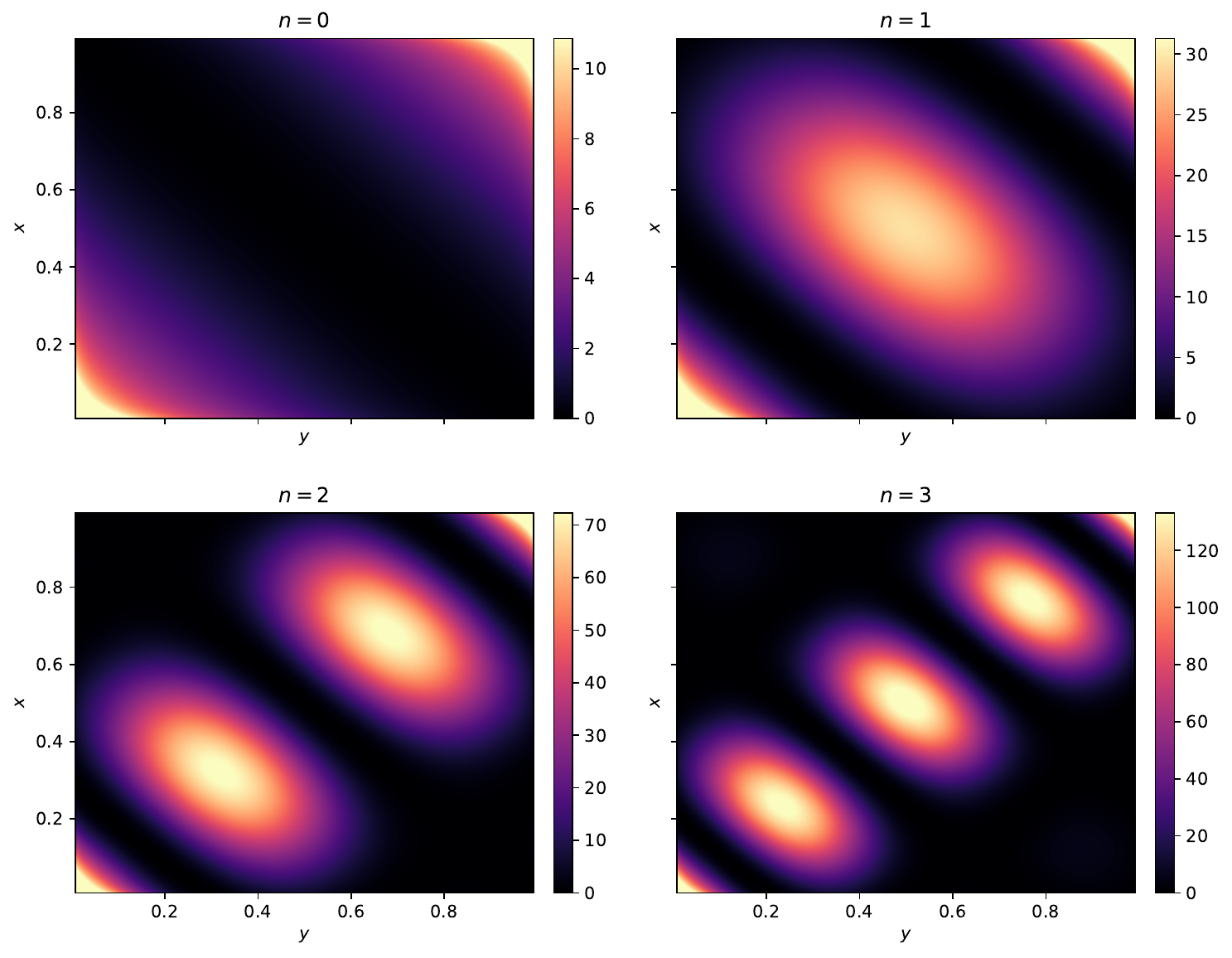}
 \caption{Exchange-symmetric Coulomb-energy kernel $\mathcal V_n(x,y)$ for the equal-mass reference family.  The kernel is nonnegative.  Each panel uses its own color scale, truncated at the 99th percentile to reveal the interior ridge structure.}
 \label{fig:kernel}
\end{figure}

The integrated interaction energy does not show how pairs of momentum fractions contribute before integration.  A natural nonnegative representation of this bilocal information is the exchange-symmetric energy kernel
\begin{equation}
 \mathcal V_n(x,y)=
 \frac{[\phi_n(x)-\phi_n(y)]^2}{2(x-y)^2}\geq0,
 \qquad
 \mathcal V_n(x,x)=\frac12[\phi_n'(x)]^2.
 \label{eq:Vkernel}
\end{equation}
Its integral is the Coulomb contribution to the invariant mass squared,
\begin{equation}
 M_{C,n}^2=\int_0^1\dd x\int_0^1\dd y\,\mathcal V_n(x,y)\geq0.
 \label{eq:VcPositive}
\end{equation}
It is strictly positive for all finite-mass states studied below, whose wave functions are nonconstant.  For fixed $0<x<1$, the diagonal value in Eq.~\eqref{eq:Vkernel} is the continuous $y\to x$ limit.  Near $x=0$ the derivative scales as $x^{\beta_1-1}$, while near $x=1$ it scales as $(1-x)^{\beta_2-1}$.  Hence the diagonal kernel diverges when $0<\beta_i<1$, whereas its two-dimensional corner integral remains finite for every $\beta_i>0$.  Fig.~\ref{fig:kernel} shows this kernel for the equal-mass reference family.  The ground-state kernel is smooth in the interior, with a near-boundary enhancement inherited from the $x^{1/2}$ behavior.  Internal wave-function nodes generate additional ridges and separated regions of large interaction energy because the amplitudes on opposite sides of a node can differ strongly and may have opposite signs.  The kernel is bilocal in longitudinal momentum fraction and is not a one-body or spatial energy density.

The analysis in this section establishes how constituent masses and excitation level shape the longitudinal meson state.  The same normalized wave functions determine the momentum distributions, the complementary localization measures, and the nonnegative bilocal kernel whose integral gives the Coulomb matrix element.  The four mass families display a continuous change from broad relativistic states, through asymmetric heavy--light configurations, to localized heavy--heavy states.

The next question is how these structural differences are reflected in the meson invariant mass squared.  I therefore turn to the expectation value of the bound-state equation, which separates the explicit quark-mass contribution from the Coulomb interaction and connects this separation with the forward EMT trace.

\section{Forward EMT structure and the diagonal GPD moment}
\label{sec:emt}

The preceding section established how the quark masses and excitation level determine the longitudinal shape of each meson state.  I now examine two related consequences for EMT observables.  The forward invariant mass squared separates into explicit quark-mass and confining-interaction contributions whose state dependence can be compared directly.  The diagonal GPD overlap then determines the part of the off-forward second moment carried by the two-body sector near $t=0$, while its nonanalytic terms reveal where nonvalence completion becomes necessary.  I begin with the forward decomposition and then turn to the elastic off-forward trajectory.

Taking the expectation value of Eq.~\eqref{eq:thooft} in a normalized eigenstate gives
\begin{equation}
 M_n^2=M_{m,n}^2+M_{C,n}^2,
 \qquad
 M_{m,n}^2=\int_0^1\dd x\,q_n(x)
 \left(\frac{\mtilde_1^2}{x}+\frac{\mtilde_2^2}{1-x}\right),
 \label{eq:massdecomp}
\end{equation}
with $M_{C,n}^2$ given by Eq.~\eqref{eq:VcPositive}.  This is the dimensionless forward mass-squared decomposition associated with the EMT trace and discussed in Refs.~\cite{Ji:2021,Freese:2023}.  Here ``forward'' means $p'=p$ and $t=(p'-p)^2=0$; no spatial decomposition is implied.

The purpose of comparing the two terms is to test whether the changes in longitudinal structure found in Sec.~\ref{sec:wf} are accompanied by a systematic change in the dynamical origin of $M_n^2$.  In particular, a heavy ground state may remain dominated by the explicit mass term, whereas increasing the excitation level generally increases the variation and nodal structure of the wave function and can enhance the Coulomb matrix element.  To compare states with very different total masses, I use the dimensionless fractions
\begin{equation}
 f_{m,n}=\frac{M_{m,n}^2}{M_n^2},
 \qquad
 f_{C,n}=\frac{M_{C,n}^2}{M_n^2},
 \qquad
 f_{m,n}+f_{C,n}=1.
 \label{eq:fractions}
\end{equation}
The decomposition in Eqs.~\eqref{eq:massdecomp}--\eqref{eq:fractions} separates the explicit quark-mass and confining-interaction contributions to the forward Hamiltonian and EMT matrix element.  It must not be identified with the DGLAP--ERBL separation used below, which separates regions of GPD support.  In particular, $f_{m,n}$ and $f_{C,n}$ do not define DGLAP or ERBL fractions and do not determine separate quark-mass and Coulomb radii.

\begin{figure}[!htbp]
 \centering
 \includegraphics[width=0.96\columnwidth]{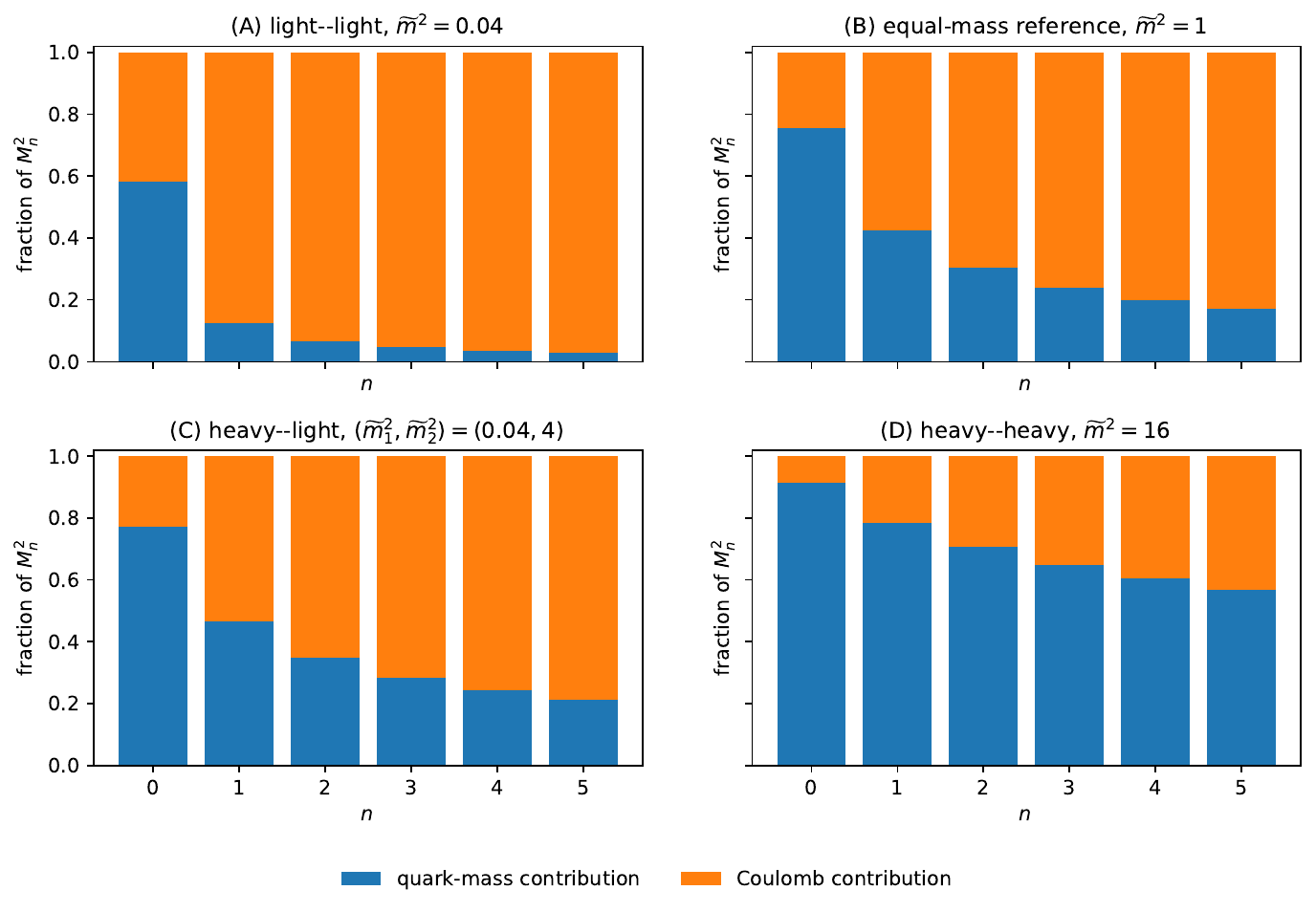}
 \caption{Fractions of $M_n^2$ carried by the explicit quark-mass and Coulomb-interaction terms.  At each displayed point, both fractions are evaluated from the same eigenstate obtained with $N_B=28$ and $N_q=2400$, so that $f_{m,n}+f_{C,n}=1$.}
 \label{fig:massdecomp}
\end{figure}

Fig.~\ref{fig:massdecomp} shows these fractions for the four mass regimes and the six lowest meson states.  The light--light ground state already has a substantial interaction contribution, $f_{C,0}\simeq0.413$, which rises to $0.875$ for the first excitation and approaches unity for higher states.  The heavy--heavy ground state is instead dominated by the explicit mass term, with $f_{C,0}\simeq0.086$, and the progression toward interaction dominance with excitation is much slower.  The equal-mass reference and heavy--light families interpolate between these limits.  Thus the balance between explicit mass and confinement is not determined by the mass parameters alone; it changes systematically with the internal structure of the eigenstate.  The Coulomb term should not be confused with a four-dimensional gluonic trace-anomaly contribution: in this model it is the energy of the instantaneous confining interaction.

\FloatBarrier

\begin{figure}[!htbp]
 \centering
 \includegraphics[width=0.92\columnwidth]{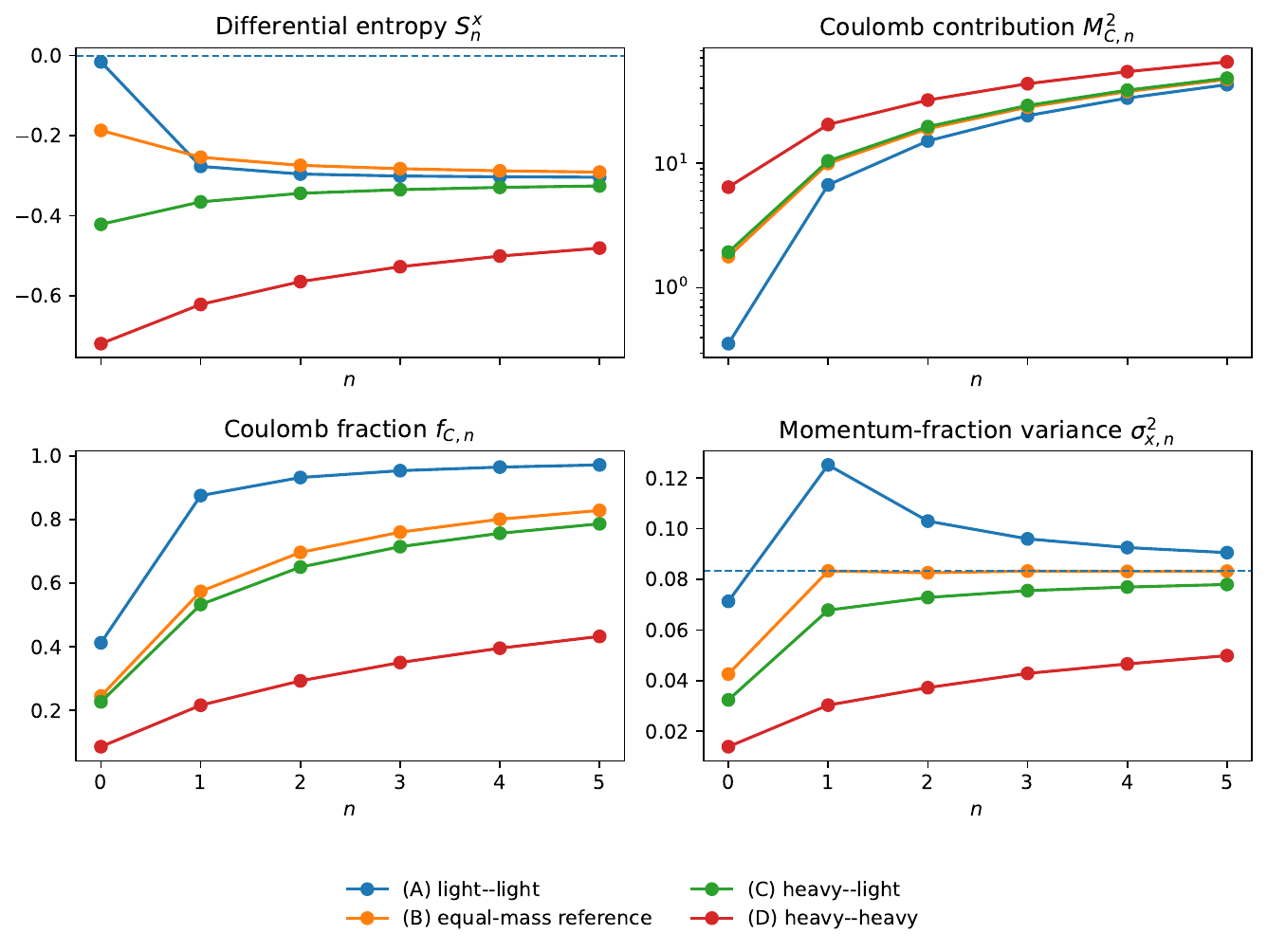}
 \caption{State-by-state comparison of longitudinal localization and Coulomb-interaction diagnostics.  All four panels use the same $N_B=28$, $N_q=2400$ solutions.  The dashed lines mark $S_n^x=0$ for the uniform distribution and $\sigma_{x,n}^2=1/12$; the Coulomb-contribution panel uses a logarithmic vertical scale.  No extrapolated quantities are combined with fixed-resolution results.}
 \label{fig:diagnostics}
\end{figure}

Fig.~\ref{fig:diagnostics} compares the differential entropy, Coulomb matrix element, Coulomb fraction, and momentum-fraction variance.  These quantities characterize the same mass- and excitation-dependent evolution from complementary viewpoints.  Light states have broad longitudinal support and become interaction-dominated rapidly with excitation.  Heavy--heavy states retain stronger localization and a large explicit-mass contribution over the same range.  The heavy--light family exhibits intermediate behavior together with a gradual redistribution toward more equal momentum sharing.  The variance and entropy need not move monotonically together because they respond differently to the separated lobes of excited-state distributions.

The heavy--heavy variance gives a particularly direct measure of this localization.  Equal constituent masses imply $\langle x\rangle_n=1/2$, so $\sigma_{x,n}^2$ measures fluctuations about equal longitudinal momentum sharing.  For the ground state, $\sigma_{x,0}^2=0.0139$ is only about $17\%$ of the uniform value $1/12$; equivalently, the rms width is $\sigma_{x,0}=0.118$, compared with $1/\sqrt{12}=0.289$ for the uniform reference.  Excitation broadens the state, but even at $n=5$ the variance is $0.0499$, only about $60\%$ of the uniform value.  The heavy mass terms therefore continue to suppress the regions near $x=0$ and $x=1$ throughout the displayed sequence.  The negative entropy values provide an independent full-distribution confirmation of this persistent localization.  The uniform density is used only as a kinematic reference of maximal entropy on $0\leq x\leq1$; distance from it measures localization in momentum fraction and should not by itself be interpreted as a measure of binding strength.

\FloatBarrier

The connection between the second GPD moment and the plus--plus EMT component is an exact local-operator identity, not an additional normalization convention.  Multiplication of the bilocal quark operator by $z$ and integration over $z$ converts the light-ray separation into a covariant derivative and produces the local twist-two operator
\begin{equation}
 T_q^{++}=\bar q\gamma^+\frac{i}{2}\overleftrightarrow D^{+}q.
 \label{eq:Tqpp}
\end{equation}
No trace-subtraction term survives in this component because $g^{++}=0$.  In the convention used here, the corresponding Mellin sum rule is~\cite{Ji:1997DVCS,Diehl:2003,Jia:2024}
\begin{equation}
 \int_{-1}^{1}\dd z\,zH_n(z,\xi,t)
 =\frac{\langle p'|T_q^{++}(0)|p\rangle}{2(P^+)^2},
 \qquad
 P^\mu=\frac{p^\mu+p^{\prime\mu}}{2}.
 \label{eq:Tppmoment}
\end{equation}
With the standard antiquark convention, the equal-mass flavor-neutral distribution at zero skewness is
\begin{equation}
 H_n(z,0,0)=
 \begin{cases}
  \phi_n^2(z), & 0<z<1,\\
  -\phi_n^2(-z), & -1<z<0.
 \end{cases}
 \label{eq:forwardGPD}
\end{equation}
Its second Mellin moment therefore satisfies
\begin{equation}
 \int_{-1}^{1}\dd z\,zH_n(z,0,0)
 =2\int_0^1\dd z\,z\phi_n^2(z)=1,
 \label{eq:forwardmoment}
\end{equation}
where reflection symmetry and wave-function normalization were used.  For the single-flavor neutral meson considered here, the flavor sum is trivial.  In $1+1$ dimensions the gauge-field contribution to $T^{++}$ vanishes, so $T_q^{++}=T^{++}$ and the quark--antiquark moment exhausts the conserved plus momentum~\cite{Ji:2021,Freese:2023,Jia:2024}.

The forward decomposition fixes how the invariant mass squared is partitioned, but it does not determine how much of the finite-$t$ EMT information is already contained in the same two-body wave function.  I address this question through the diagonal GPD contribution supported in the DGLAP region~\cite{Burkardt:2000,Jia:2024}.  For equal constituent masses, the eigenfunctions may be chosen with definite reflection parity, $\phi_n(1-x)=(-1)^n\phi_n(x)$, so that $q_n(1-x)=q_n(x)$.

For $0\leq\xi<1$, write the external plus momenta as $p^+=(1+\xi)P^+$ and $p^{\prime+}=(1-\xi)P^+$.  In the quark region $z>\xi$, the active constituent carries
\begin{equation}
 k_{\mathrm{in}}^+=(z+\xi)P^+,
 \qquad
 k_{\mathrm{out}}^+=(z-\xi)P^+,
 \label{eq:activeplus}
\end{equation}
so its fractions of the incoming and outgoing meson momenta are
\begin{equation}
 x_{\mathrm{in}}=\frac{z+\xi}{1+\xi},
 \qquad
 x_{\mathrm{out}}=\frac{z-\xi}{1-\xi}.
 \label{eq:overlapfractions}
\end{equation}
The spectator retains the same plus momentum $(1-z)P^+$ on both sides.  The diagonal two-body overlap is consequently
\begin{equation}
 H_n^{\mathrm D}(z,\xi)=
 \phi_n\!\left(\frac{z-\xi}{1-\xi}\right)
 \phi_n\!\left(\frac{z+\xi}{1+\xi}\right),
 \qquad z>\xi,
 \label{eq:HDoverlap}
\end{equation}
as derived in the light-front overlap representations of Refs.~\cite{Burkardt:2000,Jia:2024}.  The condition $z>\xi$ ensures that both active-parton plus momenta are positive.  In the antiquark region $z<-\xi$, reflection symmetry together with the conventional minus sign of the antiquark GPD makes $zH_n(z,\xi)$ equal to its quark-region counterpart.  The two diagonal regions therefore combine to give
\begin{equation}
 \mathcal M_{2,n}^{\mathrm D}(\xi)
 =2\int_\xi^1\dd z\,z\,
 \phi_n\!\left(\frac{z-\xi}{1-\xi}\right)
 \phi_n\!\left(\frac{z+\xi}{1+\xi}\right),
 \label{eq:MDxi}
\end{equation}
while negative skewness follows by interchanging the external states.  The superscript $\mathrm D$ emphasizes that only $|z|>|\xi|$ is included.  For $|z|<|\xi|$, one active-parton momentum changes sign and the bilocal operator creates or annihilates a $q\bar q$ pair rather than connecting the same two-body sector.  This ERBL contribution is therefore non-diagonal in the light-front representation and is required before Eq.~\eqref{eq:MDxi} can be identified with the complete second moment of the local EMT operator~\cite{Burkardt:2000,Jia:2024}.

In two dimensions, skewness and invariant momentum transfer are not independent because there is no transverse momentum transfer.  The origin of this relation is most transparent in an elastic Breit frame.  In Eqs.~\eqref{eq:rapiditymomenta}--\eqref{eq:telastic}, momentum components are measured in units of $\sqrt{\lambda}$.  Introducing the rapidity $\eta$ through
\begin{equation}
 p^+=\frac{M_n}{\sqrt2}e^{\eta},
 \qquad
 p^{\prime +}=\frac{M_n}{\sqrt2}e^{-\eta},
 \label{eq:rapiditymomenta}
\end{equation}
with the minus components fixed by the on-shell conditions, the symmetric skewness is
\begin{equation}
 \xi=\frac{p^+-p^{\prime +}}{p^++p^{\prime +}}
 =\tanh\eta.
 \label{eq:xirapidity}
\end{equation}
Burkardt parametrized the same elastic trajectory by the asymmetric skewness variable $\zeta$~\cite{Burkardt:2000}.  The symmetric convention used in Ref.~\cite{Jia:2024} is related by $\zeta=2\xi/(1+\xi)$.  Denoting this variable by $b\equiv\zeta$ gives
\begin{equation}
 b=\frac{2\xi}{1+\xi}=1-e^{-2\eta},
 \qquad
 \xi=\frac{b}{2-b},
 \qquad
 1-b=e^{-2\eta}.
 \label{eq:bkin}
\end{equation}
The invariant momentum transfer along the same on-shell trajectory is therefore
\begin{equation}
 t=-4M_n^2\sinh^2\eta
   =-\frac{M_n^2b^2}{1-b}.
 \label{eq:telastic}
\end{equation}
Here $t\equiv t_{\mathrm{phys}}/\lambda$ follows the convention in Eq.~\eqref{eq:units}.  The rapidity parametrization establishes the physical relation among $\eta$, $\xi$, $b$, and $t$; the fixed-interval change of variables introduced below serves the separate purpose of expanding the overlap integral.

All expansions below are performed on the spacelike elastic branch, $t\leq0$.  If the complete form factor is known as an analytic function of the Euclidean variable $Q^2=-t>0$, its physical timelike boundary value is obtained by the continuation $Q^2\to -t-i0$.  Above a multiparticle threshold this prescription selects the physical side of the branch cut and generates an imaginary part.  Stable states appear as real-axis poles, whereas unstable resonances correspond to poles on an unphysical sheet.  In the strict large-$N_c$ 't~Hooft model the mesons are stable at leading order, so the timelike singularities are zero-width poles; finite widths arise only beyond the leading large-$N_c$ limit.  No such continuation of the truncated near-forward series is attempted here, because a Taylor expansion cannot be continued reliably through its nearest pole or threshold.

To expand around the forward point, it is useful to remove the $\xi$ dependence of the integration interval.  I introduce
\begin{equation}
 x=\frac{z-\xi}{1-\xi},
 \qquad
 z=\xi+(1-\xi)x,
 \label{eq:varchange}
\end{equation}
which maps $\xi\le z\le1$ onto $0\le x\le1$.  The first wave-function argument becomes $x$, while the second becomes
\begin{equation}
 \frac{z+\xi}{1+\xi}=b+(1-b)x=x+b(1-x).
 \label{eq:shiftedarg}
\end{equation}
Using $\dd z=(1-\xi)\dd x$, the diagonal moment takes the exact form
\begin{equation}
 \MD(b)=
 \frac{4(1-b)}{(2-b)^2}
 \int_0^1\dd x\,
 \bigl[b+2(1-b)x\bigr]\phi_n(x)
 \phi_n\!\bigl(x+b(1-x)\bigr).
 \label{eq:MDexact}
\end{equation}
This representation separates the kinematic prefactor from the displacement of the second wave function and makes the expansion in $b$ transparent.

At $b=0$, normalization and reflection symmetry give
\begin{equation}
 \MD(0)=2\int_0^1x\phi_n^2(x)\,\dd x=1.
 \label{eq:MDnorm}
\end{equation}
The first derivative requires contributions from both the prefactor and the shifted wave function.  At fixed $0<x<1$, the shifted wave function has the small-$b$ expansion
\begin{equation}
 \phi_n\!\bigl(x+b(1-x)\bigr)
 =\phi_n(x)+b(1-x)\phi_n'(x)
 +\frac{b^2}{2}(1-x)^2\phi_n''(x)+\cdots,
 \label{eq:phiexpand}
\end{equation}
and
\begin{equation}
 \frac{4(1-b)}{(2-b)^2}\bigl[b+2(1-b)x\bigr]
 =2x+b(1-2x)-\frac{b^2}{2}x+O(b^3).
 \label{eq:kinexpand}
\end{equation}
Combining the term linear in $b$ from the prefactor in Eq.~\eqref{eq:kinexpand} with the linear displacement of the wave function in Eq.~\eqref{eq:phiexpand}, the first derivative at the forward point is
\begin{equation}
 \left.\frac{\dd\MD}{\dd b}\right|_{b=0}
 =\int_0^1\dd x\,
 \left[(1-2x)\phi_n^2(x)
 +2x(1-x)\phi_n(x)\phi_n'(x)\right].
 \label{eq:linearraw}
\end{equation}
The first term in Eq.~\eqref{eq:linearraw} comes from the kinematic weight, while the second comes from shifting the second wave-function argument.  Since the latter can be written as $\int_0^1x(1-x)\,\dd[\phi_n^2(x)]/\dd x\,\dd x$, integration by parts cancels the first term exactly.  The boundary term vanishes because $\phi_n(0)=\phi_n(1)=0$, and hence $\partial_b\MD(0)=0$.  The absence of a linear skewness term is therefore an exact consequence of the overlap normalization and boundary behavior, not a numerical observation.

The first surviving coefficient is quadratic.  Expanding Eq.~\eqref{eq:MDexact} through $b^2$ and integrating by parts first gives a weighted term $-\int_0^1\dd x\,x(1-x)^2[\phi_n'(x)]^2$.  Reflection symmetry converts this integral into one half of $-\int_0^1\dd x\,x(1-x)[\phi_n'(x)]^2$, yielding
\begin{equation}
 A_{2,n}^{\mathrm D}=
 -\frac14-\frac12\int_0^1\dd x\,
 x(1-x)[\phi_n'(x)]^2.
 \label{eq:A2D}
\end{equation}
The constant $-1/4$ is kinematic, whereas the second term is a weighted gradient norm of the wave function.  The factor $x(1-x)$ vanishes at the boundaries, but the mass-dependent singular behavior of $\phi_n'(x)$ can still make the near-boundary regions quantitatively important.  Increasing the excitation level does not make $A_{2,n}^{\mathrm D}$ more negative merely because the wave function contains more nodes.  Rather, the additional sign changes and separated lobes are generally accompanied by larger gradients between neighboring regions, thereby increasing
$\int_0^1\dd x\,x(1-x)[\phi_n'(x)]^2$ and making the diagonal coefficient more negative.

Equation~\eqref{eq:A2D} is the coefficient of the regular $b^2$ term.  The third-order interior expansion requires a matched prescription for sufficiently small boundary exponent.  Writing the coefficient of $b^3$ in the fixed-$x$ expansion as $\mathcal I_3(x)$, the boundary behavior $\phi_n(x)\sim C_nx^\beta$ gives
\[
 \mathcal I_3(x)\sim\frac{C_n^2}{6}\,\beta(\beta-1)(2\beta-1)x^{2\beta-2},
\]
which is not separately integrable for $0<\beta<1/2$.  The problem is nonuniformity rather than a divergence of the exact overlap: for $\phi_n(x)\sim C_nx^\beta$, the $k$th fixed-$x$ Taylor term is parametrically proportional to $(b/x)^k$.  The expansion is therefore valid in the interior, where $b\ll x$, but fails in the boundary layer $x=O(b)$ because the displacement $x\to x+b(1-x)$ is then comparable with $x$ itself.

To match the two descriptions, introduce an arbitrary intermediate scale $b\ll\epsilon\ll1$ and split the integral into $0<x<\epsilon$ and $\epsilon<x<1$.  The interior part is expanded at fixed $x$.  In the boundary part one instead sets $x=bu$ and keeps the shifted boundary power unexpanded.  A fixed small $\delta$, with $\epsilon\ll\delta\ll1$, only marks a range in which the boundary asymptotics is valid.  The leading singular interior term is
\[
 b^3\int_\epsilon^\delta\dd x\,\mathcal I_3(x)
 \supset
 -\frac{C_n^2\beta(\beta-1)}{6}\,b^3\epsilon^{2\beta-1}.
\]
The large-$u$ expansion of the $x=bu$ boundary integral supplies the opposite term, $+C_n^2\beta(\beta-1)b^3\epsilon^{2\beta-1}/6$.  Thus the dependence on the artificial matching scale cancels between the two regions, as it must; $\epsilon$ introduces no physical scale.  The finite, $\epsilon$-independent regular contribution can then be integrated by parts, giving
\begin{align}
 [b^3] \,\MD(b)\big|_{\rm reg}
 &=-\frac14-\int_0^1\dd x\,x(1-x)^2[\phi_n'(x)]^2
 \notag\\
 &=-\frac14-\frac12\int_0^1\dd x\,x(1-x)[\phi_n'(x)]^2
 \notag\\
 &=A_{2,n}^{\mathrm D}.
 \label{eq:A3D}
\end{align}
The factor $1/2$ in the second line follows explicitly from reflection symmetry.  Under $x\to1-x$, the first weighted integral becomes $\int_0^1\dd x\,x^2(1-x)[\phi_n'(x)]^2$; averaging the two equal representations gives one half of the symmetric weight $x(1-x)$.  Thus the regular $b^2$ and $b^3$ terms have the same coefficient, as summarized below.

The state dependence of the regular coefficient for the three equal-mass families is shown in Table~\ref{tab:A2D}; the increasingly negative values quantify the growth of the weighted gradient norm with excitation.

\FloatBarrier
\begin{table}[H]
\caption{Diagonal coefficient $A_{2,n}^{\mathrm D}$ for the equal-mass benchmark families.  Parentheses give the numerical uncertainty in the final displayed digits.}
\label{tab:A2D}
\begin{ruledtabular}
\begin{tabular}{lrrrr}
Family & $n=0$ & $n=1$ & $n=2$ & $n=3$ \\
\hline
light--light & $-0.3080(4)$ & $-1.5140(16)$ & $-4.0126(31)$ & $-7.979(17)$ \\
equal-mass reference & $-0.59010(18)$ & $-2.6442(14)$ & $-6.3445(46)$ & $-11.691(11)$ \\
heavy--heavy & $-2.1119(22)$ & $-8.603(15)$ & $-18.093(40)$ & $-30.156(82)$ \\
\end{tabular}
\end{ruledtabular}
\end{table}

As an independent check, expanding the exact elastic relation in Eq.~\eqref{eq:telastic} gives
\begin{equation}
 -\frac{t}{M_n^2}=\frac{b^2}{1-b}=b^2+b^3+O(b^4).
 \label{eq:tseries}
\end{equation}
Consequently, a contribution analytic and linear in $t$ along the elastic trajectory necessarily produces equal regular $b^2$ and $b^3$ coefficients.  This kinematic observation confirms, but does not replace, the direct overlap derivation and does not constrain the nonanalytic boundary terms.

The regular part of the matched expansion can therefore be written as
\begin{equation}
 \MD(b)=1+A_{2,n}^{\mathrm D}(b^2+b^3)+R_{\beta,n}(b),
 \label{eq:MDlow}
\end{equation}
where $R_{\beta,n}(b)$ collects nonanalytic boundary contributions not contained in the regular Taylor coefficients.  For $0<\beta<1/2$, the boundary layer appears at first sight to allow a term of order $b^{2+2\beta}$, which would precede the regular $b^3$ contribution.  Its finite coefficient is
\begin{equation}
 J_\beta=\operatorname{FP}\!\int_0^\infty\dd u\,
 (1+2u)u^\beta(1+u)^\beta=0.
 \label{eq:Jbeta}
\end{equation}
Here $\operatorname{FP}$ denotes the Hadamard finite part obtained after subtracting the large-$u$ powers already matched to the interior expansion.  The cancellation follows from
\[
 (1+2u)u^\beta(1+u)^\beta
 =\frac{1}{\beta+1}\frac{\dd}{\dd u}
 \left[u^{\beta+1}(1+u)^{\beta+1}\right],
\]
because, after the large-$u$ powers matched to the interior region are subtracted, the primitive has no finite constant term at infinity.

The vanishing finite part removes the nominal $b^{2+2\beta}$ contribution from the product of the two leading singular boundary branches.  The remaining endpoint corrections arise from higher terms in the generalized Frobenius expansion and from matched endpoint--bulk contributions.  A complete classification of this general-$\beta$ remainder requires the full inhomogeneous endpoint problem for the diagonal overlap.  Since the same hierarchy supplies the DGLAP-side input to the DGLAP--ERBL matching, it is developed together with that matching in Part~II rather than assumed here.

The regular coefficients in Eq.~\eqref{eq:MDlow} are completely fixed by the diagonal overlap.  The quantity $-A_{2,n}^{\mathrm D}/M_n^2$ is therefore the slope inferred from that overlap along the physical trajectory.  Its identification with the slope of the complete EMT form factor additionally requires establishing that the ERBL contribution has no term linear in $t$, which is addressed in Part~II.  At the equal-mass reference point $\mtilde^2=1$, for which $\beta=1/2$, the leading nonanalytic diagonal remainder can nevertheless be derived explicitly within the present analysis.

Let $\mathcal L_{\mathrm{end}}$ denote the leading endpoint, or indicial, operator obtained from Eq.~\eqref{eq:thooft} by retaining the explicit $1/x$ mass term and the singular integral as $x\to0$.  On a trial power $x^\alpha$, both retained terms scale as $x^{\alpha-1}$, whereas the eigenvalue term and other regular contributions begin at order $x^\alpha$ and do not enter the leading indicial balance.  This is the integral-equation analogue of the Frobenius construction for a singular differential equation.  Its action is~\cite{tHooft:1974}
\begin{equation}
 \mathcal L_{\mathrm{end}}[x^\alpha]\supset F(\alpha)x^{\alpha-1},
 \qquad
 F(\alpha)=\mtilde^2-1+\pi\alpha\cot(\pi\alpha).
 \label{eq:Falpha}
\end{equation}
At $\mtilde^2=1$, both $\alpha=1/2$ and $\alpha=3/2$ are roots.  The required $x^{3/2}$ correction is resonant because $3/2$ is itself a root of the homogeneous boundary equation.  Since
\begin{equation}
 \mathcal L_{\mathrm{end}}[x^\alpha\ln x]\supset
 \bigl[F(\alpha)\ln x+F'(\alpha)\bigr]x^{\alpha-1},
 \label{eq:logoperator}
\end{equation}
the terms of order $x^{1/2}$ obey
\[
 \mathcal L_{\mathrm{end}}\!\left[D_nx^{3/2}\ln x\right]
 \supset D_nF'(3/2)x^{1/2},
 \qquad
 M_n^2\phi_n(x)\supset M_n^2C_nx^{1/2}.
\]
Thus $D_nF'(3/2)=M_n^2C_n$.  With $F'(3/2)=-3\pi^2/2$, this gives
\begin{equation}
 \phi_n(x)=C_nx^{1/2}+D_nx^{3/2}\ln x+K_nx^{3/2}+\cdots,
 \qquad
 D_n=-\frac{2M_n^2}{3\pi^2}\,C_n.
 \label{eq:endexp}
\end{equation}
To display how this term enters the overlap, write the integrand of Eq.~\eqref{eq:MDexact} as $\sum_{k\geq0}b^k\mathcal I_k(x)$.  The only fourth-order terms capable of producing $\ln x/x$ are contained in
\begin{equation}
 \mathcal I_4(x)\supset
 \frac{(1-2x)(1-x)^3}{6}\,\phi_n\phi_n'''
 +\frac{x(1-x)^4}{12}\,\phi_n\phi_n''''.
 \label{eq:F4origin}
\end{equation}
Substituting Eq.~\eqref{eq:endexp}, the $C_nD_n$ part of these terms is
\begin{equation}
 \left.\mathcal I_4(x)\right|_{C_nD_n}
 =-\frac{C_nD_n}{32}\frac{\ln x}{x}
  -\frac{C_nD_n}{24}\frac1x+O(\ln x)
 \qquad (b\ll x\ll1).
 \label{eq:F4singular}
\end{equation}
The complete $\mathcal I_4(x)$ also contains stronger power terms, beginning with
$-C_n^2/(64x^2)$.  These power singularities belong to the matched boundary contribution and do not alter the matching-scale-independent double logarithm.  The logarithmic term itself gives, over the overlap region $b<x<\delta$ with $b\ll\delta\ll1$,
\begin{equation}
 b^4\int_b^\delta\dd x\,
 \left[-\frac{C_nD_n}{32}\frac{\ln x}{x}\right]
 =\frac{C_nD_n}{64}b^4
 \left[\ln^2\frac1b-\ln^2\frac1\delta\right].
 \label{eq:Dlogmatch}
\end{equation}
The remaining terms in $\mathcal I_4(x)$ generate powerlike, single-logarithmic, and further $\delta$-dependent contributions.  After the interior and boundary regions are combined, the powerlike and matching-scale-dependent pieces cancel, whereas the coefficient of $b^4\ln^2(1/b)$ remains unchanged.  The surviving double logarithm is
\begin{equation}
 \MD(b)\supset
 \frac{C_nD_n}{64}\,b^4\ln^2\frac1b,
 \qquad \beta=\frac12.
 \label{eq:Dlog}
\end{equation}
Since $C_nD_n=-(2M_n^2/3\pi^2)C_n^2<0$, the double-logarithmic coefficient is negative.  Its nonanalytic functional form is $t^2\ln^2(-t)$, equivalently $t^2\ln^2[M_n^2/(-t)]$, up to an overall multiplicative coefficient, terms with fewer logarithms, and analytic terms.  In the massive theory, the lowest physical $t$-channel singularity lies at positive $t$, leaving a finite neighborhood of $t=0$ in which the complete local EMT form factor is analytic.  The nonanalyticity is therefore not an inconsistency of the wave function; it identifies the support-dependent term compensated by the ERBL contribution in the complete moment.

This section has connected the longitudinal wave-function structure to two complementary aspects of the EMT.  At $t=0$, the bound-state equation separates the invariant mass squared into explicit quark-mass and Coulomb-interaction contributions and shows how their balance changes with mass and excitation.  Along the elastic trajectory, the diagonal GPD overlap has an exact normalization and a controlled near-forward expansion determined by the same wave function.  At the same time, the logarithmic boundary term identifies the nonanalytic component for which the diagonal two-body description must be completed by the nonvalence region.  This boundary obstruction is the starting point for the ERBL analysis in Part~II.

\section{Conclusions}
\label{sec:conclusions}

I have examined how much information about the meson EMT is already contained in the leading light-front wave function of the large-$N_c$ 't~Hooft model.  The comparison of light--light, equal-mass reference, heavy--light, and heavy--heavy systems shows how constituent masses and excitation level reorganize longitudinal momentum sharing.  Light states remain broadly distributed in $x$, heavy--light states retain a pronounced asymmetry at low excitation, and heavy--heavy ground states are concentrated near equal sharing.  The variance and differential entropy provide complementary descriptions of this behavior: the former measures the second central moment, whereas the latter responds to the full multi-lobed distribution generated by internal wave-function nodes.

The same normalized eigenstates determine the forward EMT trace decomposition of $M_n^2$.  The numerical comparison shows that the balance between explicit quark mass and confinement is not fixed by the mass parameters alone.  Increasing excitation level is accompanied by greater wave-function variation and a larger Coulomb contribution, while heavy ground states retain a dominant explicit-mass component.  A principal result of the present state-by-state analysis is therefore the direct correlation between longitudinal wave-function structure and the dynamical origin of the invariant mass squared.

The symmetric bilocal kernel $\mathcal V_n(x,y)$ adds information that is lost in the integrated Coulomb matrix element.  It is nonnegative and shows directly which pairs of momentum fractions carry the interaction energy.  Internal wave-function nodes create additional ridges and separated regions of large kernel strength because the wave function changes rapidly across the corresponding parts of the interval.  This is a bilocal longitudinal representation of the confining contribution, not a positive one-body density and not a spatial energy density.

The near-forward diagonal GPD analysis supplies a second set of results.  I have shown explicitly why the term linear in the asymmetric skewness variable vanishes and expressed the first nonzero coefficient as the sum of a universal kinematic term and a weighted norm of $\phi_n'(x)$.  At the equal-mass reference point $\mtilde^2=1$, the resonance of boundary powers produces the $x^{3/2}\ln x$ term, which in turn generates a $b^4\ln^2(1/b)$ contribution to the diagonal second moment.  This provides a concrete example of how near-boundary behavior of a light-front wave function controls the analytic structure of an off-forward observable.

The 't~Hooft equation, its leading boundary exponents, the forward sum rules, and the complete GPD construction are established ingredients.  The present contribution applies these ingredients within a unified state-by-state analysis of the wave-function shape, longitudinal localization, bilocal Coulomb kernel, forward EMT decomposition, and near-forward diagonal moment.  Within this common comparison, the weighted-gradient expression, the boundary-resonance analysis, and the resulting diagonal nonanalyticity identify in a particularly transparent way both the information carried by the two-body overlap and the information it cannot supply alone.

The model remains purely longitudinal and cannot reproduce the transverse gravitational densities of a four-dimensional hadron.  Its strength is instead the exact control it offers over the relation between confinement, GPD support, and EMT matrix elements.  The diagonal analysis isolates the wave-function contribution to the near-forward moment and, through its boundary-generated nonanalyticity, shows precisely why the ERBL region is indispensable for the complete local form factor.  The companion Part II paper constructs this ERBL completion, analyzes the cancellation of the support-dependent nonanalyticities, and develops the physical EMT slope, curvature, longitudinal trace structure, and finite-$t$ pole content.

\appendix
\section{Numerical convergence and independent spectrum check}
\label{app:numerics}

The analytic inputs to the calculation are the bound-state equation, the mass-dependent boundary exponents, and the symmetric quadratic form of the Coulomb operator.  The spectra, wave functions, distributions, matrix elements, and figures reported here are obtained for the mass choices in Eq.~\eqref{eq:massfamilies}.

For each family, the basis functions in Eq.~\eqref{eq:jacobibasis} are evaluated with the appropriate pair $(\beta_1,\beta_2)$.  The matrices $H$ and $A$ are then constructed, Eq.~\eqref{eq:generalizedEigen} is solved as a real symmetric generalized eigenproblem, and the eigenvectors are ordered and normalized according to Eq.~\eqref{eq:Anorm}.  All observables are evaluated from the same normalized states.  The mapping $x=(1-\cos\theta)/2$ is used in every one- and two-dimensional integral so that the rapidly varying boundary regions receive enhanced resolution.

The light--light spectrum is independently checked without factoring the boundary power into the basis.  With
\begin{equation}
 x=\frac{1-\cos\theta}{2},
 \qquad
 \phi_n(\theta)=\sum_{k=1}^{N_s}a_k^{(n)}\sin(k\theta),
 \label{eq:sinebasis}
\end{equation}
evaluating the subtracted Coulomb quadratic form on the sine modes and combining it with the mass term at $\mtilde^2=1$ cancels the off-diagonal pieces.  The equal-mass Hamiltonian may therefore be written relative to this exactly diagonal reference form as
\begin{align}
 H_{mk}^{(s)}&=\frac{\pi^2}{2}\,k\,\delta_{mk}
 +(\mtilde^2-1)Q_{mk},
 &Q_{mk}&=2\int_0^\pi\dd\theta\,
 \frac{\sin(m\theta)\sin(k\theta)}{\sin\theta},
 \notag\\
 A_{mk}^{(s)}&=\frac12\int_0^\pi\dd\theta\,
 \sin\theta\sin(m\theta)\sin(k\theta).
 \label{eq:sinematrices}
\end{align}
This gives a variational problem with a basis and boundary treatment independent of Eq.~\eqref{eq:jacobibasis}.  Since $\phi_n(\theta)\sim\theta^{2\beta}$, the nonvanishing large-$k$ sine coefficients behave as $a_k^{(n)}\sim k^{-1-2\beta}$.  The diagonal reference form in Eq.~\eqref{eq:sinematrices} is proportional to $k$, rather than $k^2$, and its omitted tail therefore scales as
\begin{equation}
 \sum_{k>N_s} k\,|a_k^{(n)}|^2=O(N_s^{-4\beta}).
 \label{eq:sinetail}
\end{equation}
The mass-correction form has the same boundary scaling: near $\theta=0$, its quadratic form contains $\int\dd\theta\,\phi_n^2(\theta)/\sin\theta\sim\int\dd\theta\,\theta^{4\beta-1}$, so the unresolved interval $\theta\lesssim N_s^{-1}$ also contributes $O(N_s^{-4\beta})$.  Because the variational coefficients are reoptimized at each $N_s$, these tail estimates motivate, but do not by themselves prove, the leading eigenvalue correction.  These two $N_s^{-4\beta}$ contributions can partially cancel.  For the four light--light states considered here, however, no cancellation of the $N_s^{-4\beta}$ term is observed: fits in which the exponent is left free in
$M_n^2(N_s)=M_n^2(\infty)+a_nN_s^{-p}+b_nN_s^{-1}$ give $p=0.4400$--$0.4460$ for $N_s\geq400$, consistent with $4\beta=0.43930$.  Moving the lower fit boundary from $N_s=200$ to $400$ changes the fitted exponent by at most $3.7\times10^{-3}$.  I therefore use the endpoint-controlled leading correction together with a subleading $N_s^{-1}$ term,
\begin{equation}
 M_n^2(N_s)=M_n^2(\infty)+a_nN_s^{-4\beta}+b_nN_s^{-1}.
 \label{eq:sineextrap}
\end{equation}
Table~\ref{tab:lightvalidation} compares this independent limit with the boundary-adapted result.  The central values differ by at most $3.4\times10^{-4}$.  Varying the sine-basis fit window and the subleading extrapolation term gives a more conservative sensitivity envelope of $2.1\times10^{-3}$.  The uncertainties in Table~\ref{tab:spectrum} are instead conservative quadrature-extrapolation uncertainties of the boundary-adapted calculation.  Thus Table~\ref{tab:lightvalidation} independently validates the light-spectrum central values, while the larger quoted uncertainties retain the separate quadrature sensitivity.

\FloatBarrier
\begin{table}[!htbp]
\caption{Independent check of the light--light spectrum.  The reference sine-basis limits use $N_s=400$--$1600$ in Eq.~\eqref{eq:sineextrap}; alternative subleading extrapolation forms give the systematic spread described in the text.}
\label{tab:lightvalidation}
\begin{ruledtabular}
\begin{tabular}{crrr}
$n$ & boundary-adapted & sine-basis limit & difference \\
\hline
0 & $0.869387$ & $0.869166$ & $-2.21\times10^{-4}$ \\
1 & $7.660451$ & $7.660116$ & $-3.35\times10^{-4}$ \\
2 & $16.197233$ & $16.197021$ & $-2.12\times10^{-4}$ \\
3 & $25.252110$ & $25.251821$ & $-2.89\times10^{-4}$ \\
\end{tabular}
\end{ruledtabular}
\end{table}

Table~\ref{tab:basischeck} summarizes the basis-size convergence scan at fixed $N_q=1600$.  The entries are the largest relative changes between $N_B=28$ and $N_B=36$ among the states for which each quantity is defined.  The light--light ground-state Coulomb fraction is the least stable displayed scalar quantity, but its basis dependence remains below $6\times10^{-4}$.

\begin{table}[!htbp]
\caption{Largest relative change between $N_B=28$ and $N_B=36$ at fixed $N_q=1600$.}
\label{tab:basischeck}
\begin{ruledtabular}
\begin{tabular}{lcc}
Quantity & Maximum relative change & Least stable state \\
\hline
$M_n^2$ & $4.7\times10^{-6}$ & light--light, $n=0$ \\
$f_{C,n}$ & $5.7\times10^{-4}$ & light--light, $n=0$ \\
$S_n^x$ & $2.5\times10^{-5}$ & equal-mass, $n=5$ \\
$\sigma_{x,n}^2$ & $8.0\times10^{-7}$ & light--light, $n=0$ \\
$A_{2,n}^{\mathrm D}$ & $9.7\times10^{-5}$ & light--light, $n=0$ \\
\end{tabular}
\end{ruledtabular}
\end{table}
\FloatBarrier

The quadrature scan uses $N_q=600,800,1100,1500,2000,$ and $2400$.  The sequences for $M_n^2$, $f_{C,n}$, and $\sigma_{x,n}^2$ are monotonic for all displayed states.  The entropy and gradient coefficient are also asymptotic over this range, with only small high-order turns for a few light excited states.  I fit the asymptotic behavior to
\begin{equation}
 X(N_q)=X_\infty+aN_q^{-p},
 \label{eq:quadfit}
\end{equation}
retain the fit beginning at $N_q=800$, and repeat it after adding the $N_q=600$ point or removing the $N_q=800$ point.  The quoted uncertainty is the largest of the shift from the $N_q=2400$ value and the changes under these fit-window variations.  Table~\ref{tab:quadcheck} gives the ground-state masses; the light families show the expected slower boundary convergence.

\begin{table}[!htbp]
\caption{Ground-state quadrature results.  The last column gives the retained extrapolated value and its numerical uncertainty.}
\label{tab:quadcheck}
\begin{ruledtabular}
\begin{tabular}{lccc}
Family & $(\beta_1,\beta_2)$ & $M_0^2(N_q=2400)$ & $M_0^2$ \\
\hline
light--light & $(0.10983,0.10983)$ & $0.86559$ & $0.8694(38)$ \\
equal-mass reference & $(0.50000,0.50000)$ & $7.27265$ & $7.2745(19)$ \\
heavy--light & $(0.10983,0.78166)$ & $8.53076$ & $8.5371(64)$ \\
heavy--heavy & $(0.93824,0.93824)$ & $74.79847$ & $74.8065(81)$ \\
\end{tabular}
\end{ruledtabular}
\end{table}
\FloatBarrier
\enlargethispage{1\baselineskip}

The $A$-metric orthonormality and the forward identity $M_n^2=M_{m,n}^2+M_{C,n}^2$ close to approximately $10^{-12}$ for every retained state at fixed $(N_B,N_q)$.  Figures~\ref{fig:massdecomp} and~\ref{fig:diagnostics} therefore use common $N_B=28$, $N_q=2400$ eigenstates.  Separate asymptotic fits of $M_n^2$, $M_{m,n}^2$, and $M_{C,n}^2$ need not preserve closure; the largest mismatch is $3.4\times10^{-2}$ for the light--light state $n=4$, and these fitted components are not used in the fraction plots.  The same fixed-resolution calculation reproduces $\mathcal M_{2,n}^{\mathrm D}(0)=1$ and $\partial_b\mathcal M_{2,n}^{\mathrm D}(0)=0$.

\renewcommand{\bibsection}{}
\par\bigskip
\noindent\rule{\textwidth}{0.4pt}
\par\medskip
\bibliography{thooft_lightfront_diagnostics_part1_review28}

\end{document}